\documentclass[a4paper,11pt]{article}
\pdfoutput=1 

\usepackage{jcappub} 

\usepackage[T1]{fontenc} 

\title{\boldmath A note on the birefringence angle estimation in CMB data analysis}

 \author[a,b]{A.~Gruppuso,}
  \author[c]{G.~Maggio,}
 \author[d,a]{D.~Molinari,}
 \author[d]{P.~Natoli}

\affiliation[a]{INAF, Istituto di Astrofisica Spaziale e Fisica Cosmica di Bologna, \\
via P.~Gobetti 101, I-40129 Bologna, Italy}
\affiliation[b]{INFN, Sezione di Bologna, Via Irnerio 46, I-40126 Bologna, Italy}
\affiliation[c]{INAF, Osservatorio Astronomico di Trieste, Via G.B. Tiepolo 11, Trieste, Italy}
\affiliation[d]{Dipartimento di Fisica e Scienze della Terra and INFN,
Universit\`a degli Studi di Ferrara, Via Saragat 1, I-44100 Ferrara, Italy}

\emailAdd{gruppuso@iasbo.inaf.it}
\emailAdd{maggio@oats.inaf.it}
\emailAdd{molinari@iasfbo.inaf.it}
\emailAdd{ntlpla@unife.it}

\abstract{Parity violating physics beyond the standard model of particle physics induces a rotation of the linear polarization of photons.
This effect, also known as cosmological birefringence (CB), can be tested with the observations of the cosmic microwave background (CMB)
anisotropies which are linearly polarized at the level of $5-10\%$. 
In particular CB produces non-null CMB cross correlations between temperature and B mode-polarization, and between E- and B-mode polarization. 
Here we study the properties of the so called D-estimators, often used to constrain such an effect. 
After deriving the framework of both frequentist and Bayesian analysis, 
we discuss the interplay between birefringence and weak-lensing, which, albeit parity conserving, modifies pre-existing TB and EB cross correlation.
}

\begin{document}

\maketitle
\flushbottom

\section{Introduction}
\label{introduction}

Polarized cosmic microwave background (CMB) data can be used to probe the cosmic birefringence effect \cite{Hinshaw:2013,Wu:2008qb,Kaufman:2013vbd}, 
i.e. the in vacuo rotation of the photon polarization direction during propagation \cite{Carroll:1989vb,Gluscevic:2010vv}.
Such an effect is naturally parameterized by an angle $\alpha$ and 
results in a mixing between Q and U Stokes parameters. 
From an observational point of view, the latter mixing produces non-null CMB cross correlations between temperature and B mode-polarization, 
and between E- and B-mode polarization\footnote{We use the so called ``cosmo'' convention for the polarization angle. 
This is typically adopted in CMB analysis, see e.g. http://wiki.cosmos.esa.int/planckpla2015/index.php/Sky\_temperature\_maps.}. 
Since these correlations are expected to be null under the parity conserving assumptions that are beneath the standard cosmological model, cosmic birefringence
represents a well suited tracer of parity violating physics.

The assumption of constant $\alpha$ defines the case of ``isotropic birefringence''.
In this case the effect on the CMB spectra is that of a rotation \cite{Lue:1998mq,Feng:2004mq}:
\begin{eqnarray}
\langle C_{\ell}^{TE,obs} \rangle &=&  \langle C_{\ell}^{TE} \rangle \cos (2 \alpha) \, ,
\label{TEobs} \\
\langle C_{\ell}^{TB,obs} \rangle &=&  \langle C_{\ell}^{TE} \rangle \sin (2 \alpha) \, ,
\label{TBobs} \\
\langle C_{\ell}^{EE, obs} \rangle &=& \langle C_{\ell}^{EE} \rangle  \cos^2 (2 \alpha) +  \langle C_{\ell}^{BB} \rangle \sin^2 (2 \alpha) \, ,
\label{EEobs} \\
\langle C_{\ell}^{BB, obs} \rangle &=&  \langle C_{\ell}^{BB}  \rangle \cos^2 (2 \alpha) +  \langle C_{\ell}^{EE} \rangle \sin^2 (2 \alpha) \, ,
\label{BBobs} \\
\langle C_{\ell}^{EB, obs} \rangle &=& {1 \over 2} \left(  \langle C_{\ell}^{EE} \rangle  - \langle C_{\ell}^{BB}  \rangle \right) \sin (4 \alpha) \, ,
\label{EBobs}
\end{eqnarray}
with $C_{\ell}^{obs}$ being the observed spectra whereas $C_{\ell}$ represent the primordial spectra, i.e. the spectra that would arise in absence of birefringence. 
Note that the primordial spectra are supposed to be parity conserving, namely $C_{\ell}^{TB}$ and $C_{\ell}^{EB}$ are vanishing. 
The symbol $\langle ... \rangle$ stands for ensemble average. Generalizations to non-constant $\alpha$ can be found e.g. in \cite{Liu:2006uh,Li:2008tma,Finelli:2008jv}.

Current constraints from CMB experiments or other astrophysical observations are all compatible with a null effect, see for instance \cite{Kaufman:2014rpa,Alighieri:2015hta,Alighieri:2016zgb}.
In CMB data analysis there are at least three ways of constraining $\alpha$:
\begin{itemize}
	\item  Monte Carlo Markov Chain (MCMC) approach. 
	In this case MCMC sampling is performed simultaneously over an extended $\Lambda$CDM model to include the birefringence angle $\alpha$, see e.g. \cite{Hinshaw:2013,Gruppuso:2015xza}.
	This approach assumes that a CMB likelihood is available and coupled to MCMC sampler. 
	Its usefulness lies in the fact that possible correlations among the parameters can be studied. However, the TB and EB spectra, which are the most 
	sensitive to parity violating mechanism, are not always included in the likelihood function because they do not add much information to the standard $\Lambda$CDM parameters or its parity conserving extensions. 
	It is possible to estimate $\alpha$ without $C_{\ell}^{TB}$ and $C_{\ell}^{EB}$ (see equations (\ref{TEobs}),(\ref{EEobs}),(\ref{BBobs})), 
	but constraints will be poorer. This is evident once Equations (\ref{TEobs})-(\ref{EBobs}) are Taylor-expanded for small $\alpha$: it turns out that only 
	Equations (\ref{TBobs}) and (\ref{EBobs}) show a linear dependence on $\alpha$ which is instead quadratic, and therefore weaker, for Equations (\ref{TEobs}),(\ref{EEobs}),(\ref{BBobs}). 
	Note also that the sign of the angle $\alpha$ cannot be determined without TB and EB.
	\item Stacking analysis. Specific pixel based estimators can be built by stacking polarization maps on temperature extrema \cite{Komatsu:2010fb,Contaldi:2015boa} or on E-mode extrema \cite{PlanckA23}.
	It is possible to show that the former are sensitive to the information contained in the TB spectrum and the latter to EB.
	\item D-estimators. These are harmonic based estimators defined through the CMB spectra by the following equations \cite{Wu:2008qb,Gruppuso:2011ci,Zhao:2015mqa,Molinari2016}:
	\begin{eqnarray}
  		D_{\ell}^{TB, obs} &=& C^{TB, obs}_{\ell} \cos (2 \beta) - C^{TE, obs}_{\ell} \sin (2 \beta)  \, , \label{DTB} \\
  		D_{\ell}^{EB, obs} &=& C^{EB, obs}_{\ell} \cos(4\beta) - \frac{1}{2} (C^{EE, obs}_{\ell} - C^{BB, obs}_{\ell}) \sin(4\beta) \, , \label{DEB}
	\end{eqnarray}
	where $\beta$ is an estimate for the birefringence angle $\alpha$. One of their most important features is that they depend explicitly on the multipole $\ell$. 
	This makes them also suitable to search for scale dependence of the birefringence effect. 
	Isotropic birefringence as described by Equations (\ref{TEobs})-(\ref{EBobs}) is completely degenerate with 
	a systematic, unknown mismatch of the global orientation of the polarimeters. 
	This systematics would clearly generate a scale independent effect. 
	On the contrary most of the birefringence models predict some form of angular dependence of $\alpha$.
	For example if the birefringence angle is proportional to the distance travelled by photons \cite{Carroll:1989vb}, then 
	one expects it to differ between small and large angular scales, since in the latter case CMB polarization is sourced at the re-ionization epoch, as opposed to 
	the former in which polarization is generated at recombination. Therefore scale dependence of $\alpha$ might be considered as a way to disentangle 
	a pure systematic effect from a fundamental physical mechanism.
	The properties of the D-estimators will be described in detail in Section \ref{description}.
\end{itemize}
Stacking and D-estimators analyses are 
sensitive to the TB and EB spectra and are typically employed fixing other cosmological parameters to the best fit model.
In the present article we focus on the D-estimators exploring in detail several remarkable properties they exhibit and 
and providing mathematical expressions that can be fruitfully used both in frequentist and Bayesian analysis.
We observe that it is convenient (from the data analysis point of view) to express all the relevant equations in terms of primordial spectra, i.e. unrotated spectra,
since in that case the covariance matrices of the D-estimators do not depend on $\alpha$ if the E and B-modes noise are equivalent\footnote{This is typically a good assumption.}.
Finally we show that weak-lensing which obviously impact CMB polarization spectra does not modify a $\chi^2$ built with the D-estimators.


This paper 
is organized as follows: in Section \ref{description} we describe the properties of the D-estimators
and provide the relevant expressions that can be used to perform the birefringence angle analysis. 
Section \ref{interplay} is devoted to the interplay between D-estimators and weak-lensing effect.
Conclusions are drawn in Section \ref{conclusion}. 


\section{D-estimators}
\label{description}

\subsection{Expectation values}
\label{expectation}

Taking the ensamble average of (\ref{DTB}) and (\ref{DEB}), one finds
\begin{eqnarray}
  	\langle D_{\ell}^{TB, obs} (\beta) \rangle &=& \langle C^{TB, obs}_{\ell} \rangle \cos (2 \beta) - \langle C^{TE, obs}_{\ell} \rangle \sin (2 \beta)  \, , \label{DTBav} \\
  	\langle D_{\ell}^{EB, obs} (\beta) \rangle &=& \langle C^{EB, obs}_{\ell} \rangle \cos(4\beta) - \frac{1}{2} (\langle C^{EE, obs}_{\ell} \rangle - \langle C^{BB, obs}_{\ell}) \rangle \sin(4\beta) \, , \label{DEBav}
\end{eqnarray}
where we have explicitly written the dependence on $\beta$ for sake of clarity.
Replacing Equations (\ref{TEobs}-\ref{EBobs}) in equations (\ref{DTBav}) and (\ref{DEBav}), after some algebra one obtains:
\begin{equation}
\langle D_{\ell}^{TB, obs} (\beta) \rangle = \langle C_{\ell}^{TE} \rangle \sin (2 (\alpha - \beta)) \, ,
\label{expDTB}
\end{equation}
\begin{equation}
\langle D_{\ell}^{EB, obs} (\beta) \rangle = \frac{1}{2} \, \left( \langle C_{\ell}^{EE} \rangle - \langle C_{\ell}^{BB} \rangle \right) \, \sin (4 (\alpha - \beta)) \, ,
\label{expDEB}
\end{equation}
where primordial $C_{\ell}^{TB}$ and $C_{\ell}^{EB}$ are set to zero\footnote{This is a work hypothesis that holds throughout the paper.}.
It is easy to see from Equations (\ref{expDTB}) and (\ref{expDEB}) that the functions $\langle D_{\ell}^{TB, obs} (\beta) \rangle$ and $\langle D_{\ell}^{EB, obs} (\beta) \rangle$ 
vanish when
\begin{equation}
\beta = \alpha \, .
\label{betaugualealpha}
\end{equation}
Therefore, looking for $\beta$ that nulls the expectation values
of the D-estimators, is equivalent to looking for the birefringence angle $\alpha$ that has rotated the primordial CMB spectra.
From a data analysis point of view such a search of $\beta$ is conveniently performed with a standard $\chi^2$ technique which 
is presented in Section \ref{chi2section}. From now on, unless otherwise specified, we always consider that Equation (\ref{betaugualealpha}) holds.

\subsection{Single realization results}

Equations (\ref{TEobs}-\ref{EBobs}) follow from
\begin{eqnarray}
a_{\ell m}^{T,obs} &=& a_{\ell m}^{T} + a_{\ell m}^{n,T} \, , \label{almT} \\
a_{\ell m}^{E,obs} &=& a_{\ell m}^{E} \cos (2 \alpha) -  a_{\ell m}^{B} \sin (2 \alpha) + a_{\ell m}^{nE} \, , \label{almE}\\
a_{\ell m}^{B,obs} &=& a_{\ell m}^{B} \cos (2 \alpha) +  a_{\ell m}^{E} \sin (2 \alpha) + a_{\ell m}^{nB} \, , \label{almB}
\end{eqnarray}
where we have explicitly written the contribution of the primordial signal (i.e. $a_{\ell m}^{T}$, $a_{\ell m}^{E}$ and $a_{\ell m}^{B}$) and of the instrumental noise 
(i.e. $a_{\ell m}^{nT}$, $a_{\ell m}^{nE}$ and $a_{\ell m}^{nB}$) present on the maps\footnote{We are neglecting the contribution of spurious astrophysical emissions.}.
From Equations (\ref{almT}-\ref{almB}) one recovers (\ref{TEobs}-\ref{EBobs}) as follows:
\begin{enumerate}
\item define the angular power spectra (APS), $C_{\ell}$, as
\begin{equation}
C_{\ell}^{XY} = \frac{1}{2 \ell +1}\sum_{\ell = -m}^m a_{\ell m}^{X} (a_{\ell m}^{Y})^{\star} - b^{nX} \, \delta_{X,Y} \, ,
\label{aps}
\end{equation}
where $X,Y$ stand for $T,E,B,nT,nE$ or $nB$, with $b^{nX} = \langle C_{\ell}^{nX,nX} \rangle$. 
Sometimes we also use the following notation: $\delta C_{\ell}^{nX,nX} = C_{\ell}^{nX,nX} - b^{nX}$.
Notice that Equation (\ref{aps}) is unbiased.
\item set to zero the primordial spectra $C_{\ell}^{TB}$ and $C_{\ell}^{EB}$; 
\item perform an average over a set of realizations. 
\end{enumerate}
Replacing in Equations (\ref{DTB}, \ref{DEB}) what found in step 1 above, one obtains the following expressions (for $\beta = \alpha$):
\begin{eqnarray}
  		D_{\ell}^{TB, obs} &=& (C^{TB}_{\ell} + C^{nT,B}_{\ell} ) + (C^{T,nB}_{\ell} + C^{nT,nB}_{\ell}) \cos (2 \alpha) +  \nonumber \\ 
		& & - (C^{T,nE}_{\ell} + C^{nT,nE}_{\ell}) \sin (2 \alpha)  \label{DTBp} \\
  		D_{\ell}^{EB, obs} &=& C^{EB}_{\ell} + (C^{E,nB}_{\ell} + C^{B,nE}_{\ell}) \cos(2\alpha) + (C^{B,nB}_{\ell} - C^{E,nE}_{\ell}) \sin(2\alpha) + \nonumber \\
		& &  + C^{nE,nB}_{\ell}  \cos(4\alpha) - \frac{1}{2} (\delta C^{nE,nE}_{\ell} - \delta C^{nB,nB}_{\ell}) \sin(4\alpha) \, . \label{DEBp}
\end{eqnarray}
As a sanity test it is easy to check that taking the ensemble average of Equations (\ref{DTBp}), (\ref{DEBp}) one finds
\begin{eqnarray}
\langle D_{\ell}^{TB, obs} \rangle &=& 0 \, , \\
\langle D_{\ell}^{EB, obs} \rangle &=& 0 \, ,
\end{eqnarray}
whatever the value of $\alpha$, which confirms the result of Section \ref{expectation}.

\subsection{Covariances}
\label{covariances}

Given the null means, the covariances of the D-estimators are
\begin{equation}
\langle D_{\ell}^{TB, obs}  D_{\ell'}^{TB, obs} \rangle \, ,
\label{covTB}
\end{equation}
\begin{equation}
\langle D_{\ell}^{EB, obs}  D_{\ell'}^{EB, obs} \rangle \, ,
\label{covEB}
\end{equation}
\begin{equation}
\langle D_{\ell}^{TB, obs}  D_{\ell'}^{EB, obs} \rangle \, ,
\label{covTBEB}
\end{equation}
\begin{equation}
\langle D_{\ell}^{EB, obs}  D_{\ell'}^{TB, obs} \rangle \, ,
\label{covEBTB}
\end{equation}
and can be computed starting from Equations (\ref{DTB})-(\ref{DEB}). 
In this case one recovers what already provided in \cite{Zhao:2015mqa} where the covariances are given in terms of observed APS. 
Plugging instead Equations (\ref{DTBp})-(\ref{DEBp}) in (\ref{covTB})-(\ref{covEBTB}) allows one to connect the covariances 
with the primordial (i.e.~unrotated) APS. 
Employing Wick's theorem (all $a_{\ell m}$ are Gaussian variables) one finds
\begin{eqnarray}
  	\langle D_{\ell}^{TB, obs}  D_{\ell'}^{TB, obs} \rangle &=&  \left[ (C_{\ell}^{T} + C_{\ell}^{nT} ) ( C_{\ell}^B + 
	C_{\ell}^{nB} \cos^2(2 \alpha) + \right. \nonumber \\
	& & \left. + C_{\ell}^{nE} \sin^2(2 \alpha) ) \right] \delta_{\ell \ell'} \frac{1}{2 \ell +1} \label{covTBnew} \\
  	\langle D_{\ell}^{EB, obs}  D_{\ell'}^{EB, obs} \rangle &=& \left[  C_{\ell}^{E} C_{\ell}^B + (C_{\ell}^{E}C_{\ell}^{nB}+C_{\ell}^{B}C_{\ell}^{nE})  \cos^2(2 \alpha) + \right. \nonumber \\
	& & \left. + (C_{\ell}^{B}C_{\ell}^{nB}+C_{\ell}^{E}C_{\ell}^{nE})  \sin^2(2 \alpha) + \right. \nonumber \\
	& & \left. + (C_{\ell}^{nE}C_{\ell}^{nB})  \cos^2(4 \alpha) + \right. \nonumber \\
	& & \left. + 2 (C_{\ell}^{nE}C_{\ell}^{nE} + C_{\ell}^{nB}C_{\ell}^{nB})  \frac{1}{4} \sin^2(4 \alpha)  \right] \delta_{\ell \ell'} \frac{1}{2 \ell +1} \, , \label{covEBnew} \\
	\langle D_{\ell}^{TB, obs}  D_{\ell'}^{EB, obs} \rangle &=& \left[ C_{\ell}^{TE} (C_{\ell}^{B} + C_{\ell}^{nB} \cos^2(2 \alpha) + C_{\ell}^{nE} \sin^2(2 \alpha))  \right] \delta_{\ell \ell'} \frac{1}{2 \ell +1} 
	\, , \label{covTBEBnew} \\
	\langle D_{\ell}^{EB, obs}  D_{\ell'}^{TB, obs} \rangle &=& \langle D_{\ell'}^{TB, obs}  D_{\ell}^{EB, obs} \rangle \, . \label{covEBTBnew} 
\end{eqnarray}
These are very useful expressions for data analysis as we will see in Section \ref{chi2section}. 
Note that if $C_{\ell}^{nE} = C_{\ell}^{nB} \equiv C_{\ell}^{nP}$ (typically a good assumption) they simplify in
\begin{eqnarray}
  	\langle D_{\ell}^{TB, obs}  D_{\ell'}^{TB, obs} \rangle &=&  \left[ (C_{\ell}^{T} + C_{\ell}^{nT}) 
	(C_{\ell}^B + C_{\ell}^{nP} ) \right] \delta_{\ell \ell'} \frac{1}{2 \ell +1} \label{covTBnew2} \\
  	\langle D_{\ell}^{EB, obs}  D_{\ell'}^{EB, obs} \rangle &=& \left[ (C_{\ell}^{E}+C_{\ell}^{nP}) (C_{\ell}^{B}+C_{\ell}^{nP})  \right] \delta_{\ell \ell'} \frac{1}{2 \ell +1} \, , \label{covEBnew2} \\
	\langle D_{\ell}^{TB, obs}  D_{\ell'}^{EB, obs} \rangle &=& \left[ C_{\ell}^{TE} (C_{\ell}^{B} + C_{\ell}^{nP})  \right] \delta_{\ell \ell'} \frac{1}{2 \ell +1} 
	\, , \label{covTBEBnew2} \\
	\langle D_{\ell}^{EB, obs}  D_{\ell'}^{TB, obs} \rangle &=& \langle D_{\ell}^{TB, obs}  D_{\ell'}^{EB, obs} \rangle \, , \label{covEBTBnew2}
\end{eqnarray}
and the dependence on $\alpha$ drops out. The latter Equations are formally identical to (\ref{covTBnew})-(\ref{covEBTBnew}) with vanishing $\alpha$. 
Also note that the $D_{\ell}^{TB, obs}$ and $D_{\ell}^{EB, obs}$ are un-correlated only in the ideal case in which the primordial B-modes are zero and noise 
can be assumed to vanish, see Equations (\ref{covTBEBnew}) and (\ref{covTBEBnew2}).

\subsection{$\chi^2$ minimization}
\label{chi2section}

The best fit angle $\alpha$ can be obtained minimizing the following $\chi^2$ expressions:
\begin{eqnarray}
\chi^2_{TB} &=& \sum_{\ell, \ell'} \, D_{\ell}^{TB, obs}  M^{TB}_{\ell \ell'} D_{\ell'}^{TB, obs} \, , \label{chiTB} \\ 
\chi^2_{EB} &=& \sum_{\ell, \ell'} \, D_{\ell}^{EB, obs}  M^{EB}_{\ell \ell'} D_{\ell'}^{EB, obs} \, , \label{chiEB}  
\end{eqnarray}
where $(M^{TB/EB}_{\ell \ell'})^{-1} = \langle D_{\ell}^{TB/EB, obs} D_{\ell'}^{TB/EB, obs} \rangle $.
One can also minimize the combination
\begin{eqnarray}
\chi^2_{TB+EB} &=& \sum_{\ell, \ell'} \,  \left( D_{\ell}^{TB, obs},D_{\ell}^{EB, obs} \right) \tilde M _{\ell \ell'} \left( D_{\ell'}^{TB, obs},D_{\ell'}^{EB, obs} \right)   \, ,
 \label{chiTB+EB}
\end{eqnarray}
with $\tilde M_{\ell \ell'}$ being implicitly defined by
\begin{equation}
\tilde M^{-1}_{\ell \ell'} =
  \left[ {\begin{array}{cc}
    \langle D_{\ell}^{TB, obs} D_{\ell'}^{TB, obs} \rangle &  \langle D_{\ell}^{TB, obs} D_{\ell'}^{EB, obs} \rangle \\       
    \langle D_{\ell}^{EB, obs} D_{\ell'}^{TB, obs} \rangle &  \langle D_{\ell}^{EB, obs} D_{\ell'}^{EB, obs} \rangle \\      
    \end{array} } \right] \, .
\end{equation}
Since $\langle D_{\ell}^{TB, obs}  D_{\ell'}^{EB, obs} \rangle =0$ if $C_{\ell}^B=0$ and for vanishing noise, 
under the latter condition
\begin{equation}
\chi^2_{TB+EB} = \chi^2_{TB} + \chi^2_{EB} \, .
\end{equation}

Equations (\ref{chiTB})-(\ref{chiTB+EB}) can be minimized following either a Bayesian or frequentist approach.
In the latter case a specific model is chosen and the matrices $M$ are built for such a reference model 
(the natural choice is $\alpha=0$, at least as long as there is no clear evidence of an $\alpha$ detection). 
The idea is then to test whether the observed value of $\alpha$ is compatible with model assumption.
Through Monte Carlo simulations an empirical posterior distribution for $\alpha$ can be obtained and used 
to derive confidence intervals.
On the contrary, in the Bayesian approach the $M$ matrices keep their dependence on $\alpha$. 
The idea here is to find the angle $\alpha$ which best describes the observations. 
Within the latter approach Monte Carlo simulations are in principle unneeded to evaluate statistical uncertainties. 
However in both approaches the $M$ matrices must be explicitly known.
Failing analytical approximations they have to be derived using Monte Carlo simulations.
To this extent the computation for the Bayesian  case is readily simplified by the following argument: 
in principle for each fixed angle $\bar \alpha$ one should simulate a large number of primordial maps, rotate them by a quantity $ 2 \bar \alpha$, 
add a random noise realization compatible with the data, apply an (unbiased) APS estimator to each of the simulated maps, build the matrix $M$,
invert it and evaluate the corresponding $\chi^2(\bar \alpha)$. This pipeline should be repeated for every $\alpha \in [\alpha_{min},\alpha_{max}]$ in order to sample 
the $\chi^2$ as a function of $\alpha$ and find its minimum. In practice one would perform $N$ sets of simulations where $N=(\alpha_{max}-\alpha_{min})/\Delta \alpha$,
with $\Delta \alpha$ being the adopted step in sampling $\alpha$.
By noticing that the dependence on $\alpha$ can be factorized as in Eqs (\ref{DTBp}), (\ref{DEBp}) it is possible to perform just one set of simulations (i.e. $N=1$) for every $\alpha$ 
and build empirically the matrix $M$ (or $\tilde M$) through Eqs. (\ref{covTB}), (\ref{covEB}), (\ref{covTBEB}) and (\ref{covEBTB}). 

\subsection{$\chi^2$ minimization in subintervals of $\ell$.}
\label{alphaspectrum}

Equations (\ref{chiTB}), (\ref{chiEB}) and (\ref{chiTB+EB}) can be minimized in subintervals of the multipole range to investigate possible scale dependences of the birefringence effect.
Note that such an investigation using the other methods listed in Section \ref{introduction} is troublesome. 
Therefore the D-estimators turn out to be the most suitable for such a kind of analysis.

Let us consider now the most ideal case in which there is no correlation between $\ell$-$\ell'$ and in which the dependence of the covariance on $\alpha$ is dropped out,
see Equations (\ref{covTBnew2}) and (\ref{covEBnew2}). Mathematically it would seem possible to minimize $\alpha$ for each single $\ell$,
\begin{eqnarray}
\chi^2_{(TB),\ell} &=& (D_{\ell}^{TB, obs})^2 / \sigma^2_{TB} \, , \label{chiTBsingleell} \\ 
\chi^2_{(EB),\ell} &=& (D_{\ell}^{EB, obs})^2 / \sigma^2_{EB} \, , \label{chiEBsingleell}  
\end{eqnarray}
where $\sigma^2_{TB/EB}$ are defined by Equations (\ref{covTBnew2}),(\ref{covEBnew2}). 
In this case, expanding Equations (\ref{chiTBsingleell}), (\ref{chiEBsingleell}) for small $\alpha$ up to second order, one finds
\begin{eqnarray}
\chi^2_{(TB),\ell} (\alpha) &=& a_{TB} \, \alpha^2 - b_{TB} \, \alpha + c_{TB} \, , \label{chiTBsingleellnew} \\ 
\chi^2_{(EB),\ell} (\alpha) &=& a_{EB} \, \alpha^2 - b_{EB} \, \alpha + c_{E B}  \, , \label{chiEBsingleellnew}  
\end{eqnarray}
where
\begin{eqnarray}
a_{TB} &=& \frac{ 4 [ (C^{TE, obs}_{\ell})^2 - (C^{TB, obs}_{\ell})^2]}{ \sigma^2_{TB}} \, ,  \,\,\,\,\,   a_{EB} = \frac{ 4 [ (\Delta^{EE, obs}_{\ell})^2 - 4 (C^{EB, obs}_{\ell})^2]}{ \sigma^2_{EB}}  \, ,\\
b_{TB} &=& \frac{ 4 C^{TE, obs}_{\ell} C^{TB, obs}_{\ell}}{ \sigma^2_{TB}} \, ,  \,\,\,\,\,  b_{EB} = \frac{ 4 \Delta^{EE, obs}_{\ell} C^{EB, obs}_{\ell}}{ \sigma^2_{EB}} \, , \\
c_{TB} &=& \frac{ (C^{TB, obs}_{\ell})^2}{ \sigma^2_{TB}} \, ,  \,\,\,\,\, c_{EB} =  \frac{ (C^{EB, obs}_{\ell})^2}{ \sigma^2_{EB}} \, ,
\end{eqnarray}
with $\Delta^{EE, obs}_{\ell} \equiv C^{EE, obs}_{\ell} - C^{BB, obs}_{\ell}$.
Of course, in the considered regime of $\alpha$, Equations (\ref{chiTBsingleellnew}),(\ref{chiEBsingleellnew}) are simply parabolic functions.
Some considerations are in order:
\begin{itemize}
\item Equations (\ref{chiTBsingleellnew}),(\ref{chiEBsingleellnew}) are convex (as it must be for a $\chi^2$ statistics) parabolic functions since the coefficients of $\alpha^2$ ($a_{TB}$ and $a_{EB}$) are always non negative. 
Only in case of noise dominated regime, these coefficients tend to zero and consequently the estimators stop being capable to constrain the birefringence angle. 
\item the vertices are different from zero only if $C^{TB, obs}_{\ell} \neq 0$ and $C^{EB, obs}_{\ell} \neq 0$.
\item for both the D-estimators the uncertainties are driven by the coefficient of the $\alpha^2$ term, namely $a_{TB}$ and $a_{EB}$. 
The larger is that coefficient, the narrower is the parabola and the smaller is the uncertainty. Conversely the smaller
is that coefficient, the wider is the parabola and therefore the larger is the uncertainty. 
In practice, considering Eq. (\ref{chiTBsingleellnew}) one sees that when $C^{TE, obs}_{\ell}$ is far from zero (even negative) then the corresponding statistical uncertainty of $\alpha$ is small. 
Similarly Eq. (\ref{chiEBsingleellnew}) shows that when $\Delta^{EE, obs}_{\ell}$ is far from zero (i.e.  $C^{EE, obs}_{\ell}$ is large) then the corresponding statistical uncertainty of $\alpha$ is small.
These simple considerations represent an handle to understand the behavior of the uncertainties at different angular scale.
\end{itemize}

\section{Interplay between birefringence and weak-lensing effect}
\label{interplay}

\subsection{Weak lensing}
\label{weaklensing}

The impact of weak-lensing on CMB spectra is computed in  \cite{Challinor:2005jy} and extended to $C_\ell^{EB}$, $C_\ell^{TB}$ in \cite{Gubitosi:2014cua}.
The lensed two-point correlation functions, i.e. $\bar \xi_{\{X,+,-\}}$, are given by:
\begin{eqnarray}
\bar\xi_X(\gamma)&=& \sum_{\ell}  \frac{2\ell +1}{4\pi} (C_\ell^{TE}-i C_\ell^{TB})\Big\{d_{20}^\ell X_{022}X_{000}+C_{gl,2}\frac{2 X_{000}'}{\sqrt{\ell(\ell+1)}}(X_{112}d_{11}^\ell+X_{132} d_{3-1}^\ell)\nonumber\\
&&\qquad +\frac{1}{2} C_{gl,2}[d_{20}^\ell(2 X_{022}'X_{000}'+X_{220}^2)+d_{-24}^\ell X_{220}X_{242}] \Big\}  \label{eq:LensedXix}  \, , \\
\bar\xi_+(\gamma)&=& \sum_\ell \frac{2\ell +1}{4\pi}(C_\ell^{EE}+ C_\ell^{BB})\Big\{d_{22}^\ell X_{022}^2+2 C_{gl,2} X_{132}X_{121}d_{31}^\ell +C_{gl,2}^2[d_{22}^\ell (X_{022}')^2\nonumber\\
&&\qquad +d_{40}^\ell X_{220}X_{242}] \Big\}  \label{eq:LensedXi+}  \, , \\
\bar\xi_-(\gamma)&=& \sum_\ell \frac{2\ell +1}{4\pi}(C_\ell^{EE}-C_\ell^{BB}-2i C_\ell^{EB})\Big\{d_{2-2}^\ell X_{022}^2+C_{gl,2} [X_{121}^2d_{1-1}^\ell+X_{132}^2d_{3-3}^\ell] \nonumber\\
&&\qquad +\frac{1}{2}C_{gl,2}^2[ 2 d_{2-2}^\ell (X_{022}')^2+d_{00}^\ell X_{220}^2+d_{4-4}^\ell X_{242}^2] \Big\} \label{eq:LensedXi-} \, ,
\end{eqnarray}
with $\gamma$ being the angle between two given points of the sphere, $d_{ss'}^\ell(\gamma)\equiv \sum_m \,_sY_{\ell m}^*(\hat n_1) \,_{s'}Y_{\ell m}(\hat n_2)$ and where $X_{ijk}$ and $C_{gl,2}$ are objects that depend on the lensing potential and geometrical factors, see \cite{Challinor:2005jy,Gubitosi:2014cua} for further details. 
Following \cite{Gubitosi:2014cua}, the power spectra appearing in the right hand side of Equations (\ref{eq:LensedXix}),(\ref{eq:LensedXi+}),(\ref{eq:LensedXi-}) are unlensed, but 
they include already the effect of birefringence, if any.
Once the lensed two-point correlation functions are computed, the lensed spectra $\bar C_\ell$ can be obtained through Equations  (\ref{eq:LensedXix}),(\ref{eq:LensedXi+}),(\ref{eq:LensedXi-}) and 
employing the orthogonality properties of $d_{ss'}^\ell$: 
\begin{eqnarray}
\bar C_\ell^{TE}-i \bar C_\ell^{TB}=2\pi \int_{-1}^1 \bar \xi_X (\gamma) d_{20}^\ell (\gamma) d\cos \gamma  \label{eq:Clfromxix}  \, , \\
\bar C_\ell^{EE}+ \bar C_\ell^{BB}=2\pi \int_{-1}^1 \bar \xi_+ (\gamma) d_{22}^\ell (\gamma) d\cos \gamma  \label{eq:Clfromxi+}  \, , \\
\bar C_\ell^{EE}-\bar C_\ell^{BB}-2 i \bar C_\ell^{EB}=2\pi \int_{-1}^1 \bar \xi_- (\gamma) d_{2-2}^\ell (\gamma) d\cos \gamma \label{eq:Clfromxi-} \, .
\end{eqnarray}
Note that since the functions between curl brackets in Equations  (\ref{eq:LensedXix}),(\ref{eq:LensedXi+}),(\ref{eq:LensedXi-}) 
are real, weak-lensing modifies $C_\ell^{TE}$,$C_\ell^{TB}$,$C_\ell^{EB}$ and $(C_\ell^{EE} \pm C_\ell^{BB})$
only {\it separately}.
This shows that the weak-lensing effect is a parity conserving phenomenon.
Considering Equations (\ref{eq:Clfromxix}) and (\ref{eq:Clfromxi-}), and separating the real and imaginary part, one can write
\begin{eqnarray}
\bar C_\ell^{TE} &=& \sum_{\ell'} C_{\ell'}^{TE} G^{(1)}_{\ell \ell'} \, , \label{TElensing} \\
\bar C_\ell^{TB} &=& \sum_{\ell'} C_{\ell'}^{TB} G^{(1)}_{\ell \ell'} \, , \label{TBlensing} \\
\bar C_\ell^{EB} &=& \sum_{\ell'} C_{\ell'}^{EB} G^{(2)}_{\ell \ell'} \, , \label{EBlensing} \\
(\bar C_\ell^{EE} - \bar C_\ell^{BB}) &=& \sum_{\ell'} ( C_{\ell'}^{EE} - C_{\ell'}^{BB} ) G^{(2)}_{\ell \ell'} \, , \label{EEmBBlensing}
\end{eqnarray}
with
\begin{eqnarray}
G^{(1)}_{\ell \ell'} &=&    \frac{2 \ell' +1}{2} \int_{-1}^{1} \, F^{(1)}_{\ell'}(\gamma) d_{2 0}^\ell(\gamma) d\cos \gamma \, , \\ 
G^{(2)}_{\ell \ell'} &=&  \frac{2 \ell' +1}{2} \int_{-1}^{1} \, F^{(2)}_{\ell'}(\gamma) d_{2 -2}^\ell(\gamma) d\cos \gamma \, ,
\end{eqnarray}
where
\begin{eqnarray}
F^{(1)}_{\ell}(\gamma) &\equiv&  \Big\{d_{20}^\ell X_{022}X_{000}+C_{gl,2}\frac{2 X_{000}'}{\sqrt{\ell(\ell+1)}}(X_{112}d_{11}^\ell+X_{132} d_{3-1}^\ell)\nonumber\\
&&\qquad +\frac{1}{2} C_{gl,2}[d_{20}^\ell(2 X_{022}'X_{000}'+X_{220}^2)+d_{-24}^\ell X_{220}X_{242}] \Big\} \nonumber \, , \\
F^{(2)}_{\ell}(\gamma) &\equiv&  \Big\{d_{2-2}^\ell X_{022}^2+C_{gl,2} [X_{121}^2d_{1-1}^\ell+X_{132}^2d_{3-3}^\ell] \nonumber\\
&&\qquad +\frac{1}{2}C_{gl,2}^2[ 2 d_{2-2}^\ell (X_{022}')^2+d_{00}^\ell X_{220}^2+d_{4-4}^\ell X_{242}^2] \Big\} \, .
\end{eqnarray}


\subsection{Minimization of the $\chi^2$ built with D-estimators}
\label{chi2}

We consider now again the D-estimators, defined in Equations (\ref{DTB}) and (\ref{DEB}). 
As done in \cite{Gubitosi:2014cua} we suppose that weak-lensing is a late time effect.
Therefore after the weak-lensing takes place, the D-estimators, denoted as $\bar D_{\ell}^{TB/EB, obs}$, can be written as
\begin{eqnarray}
	\bar D_{\ell}^{TB, obs} &=& \bar C^{TB, obs}_{\ell} \cos (2 \beta) - \bar C^{TE, obs}_{\ell} \sin (2 \beta) = G^{(1)}_{\ell \ell'} D_{\ell'}^{TB, obs} \, , \label{DTBafterlensing} \\
  	\bar D_{\ell}^{EB, obs} &=& \bar C^{EB, obs}_{\ell} \cos(4\beta) - \frac{1}{2} (\bar C^{EE, obs}_{\ell} - \bar C^{BB, obs}_{\ell}) \sin(4\beta) = G^{(2)}_{\ell \ell'} D_{\ell'}^{EB, obs}  \, , \label{DEBafterlensing}
\end{eqnarray}
where the sum over $\ell'$ is understood and where Equations (\ref{TElensing})-(\ref{EEmBBlensing}) have been used.
For sake of simplicity we omit now the label $\ell$, and adopt a matrix-vector formalism.
The $\chi^2$ built with $\bar D$, for both $TB$ and $EB$, is then given by
\begin{eqnarray}
\chi^2_{\bar D} & = & \bar D^T (\langle \bar D \bar D^{T} \rangle)^{-1}  \bar D = D G^T (\langle G D D^T G^T  \rangle)^{-1} G D   \nonumber \\
	                  & = & D G^T (G^T)^{-1} (\langle D D^T \rangle)^{-1} G^{-1} G D  \nonumber \\
	                  & = & D (\langle D D^T \rangle)^{-1} D = \chi^2_{D} 
	                  \label{chiquadroinvariant} \, .
\end{eqnarray}
Equation (\ref{chiquadroinvariant}) shows that the function $\chi^2(\alpha)$ is invariant under the weak-lensing effect.
We stress that not only the minimum but the whole shape does not depend on the fact that weak-lensing took place or not.
Notice how this follows from the structure of the weak-lensing kernel and in particular its parity conserving nature as highlighted above.

We can follow two ways to estimate $\alpha$: (a) we can take the observed APS (that do include the weak-lensing) 
and build a covariance matrix {\it with} the inclusion of the weak-lensing or (b) we can de-lens the observed spectra and build the covariance
matrix with APS that do not include the weak-lensing effect. Of course both approaches assume that the lensing potential is known. 

\subsection{Mismatch in the weak-lensing matrix}
\label{mismatch}

In fact the lensing potential would be known with a certain degree of uncertainty. 
We suppose that the weak-lensing matrix adopted, $\bar G$, differ from the real weak-lensing matrix $G$ for a term proportional to a quantity $\epsilon$ 
\begin{equation}
\bar G = G \left( 1 + \epsilon \right) 
\, .
\label{barGminusG}
\end{equation}
In this case the ``wrong'' $\bar \chi^2$ statistics will be given by
\begin{equation}
\bar \chi^2 = D \bar G^T (\langle G D D^T G^T  \rangle)^{-1} \bar G D 
\, .
\label{chi2bar}
\end{equation}
Replacing Equation (\ref{barGminusG}) in Equation (\ref{chi2bar}), one finds
\begin{eqnarray}
\bar \chi^2_{\bar D} & = & D \bar G^T (G^T)^{-1} (\langle D D^T \rangle)^{-1} G^{-1} \bar G D  \nonumber \\
	                  & = & D G^T (1 + \epsilon) (G^T)^{-1} (\langle D D^T \rangle)^{-1} G^{-1} (1+\epsilon) G D  \nonumber \\
	                  & \simeq & D (\langle D D^T \rangle)^{-1} D (1 + 2 \, \epsilon) = \chi^2_{D} (1 + 2 \, \epsilon)
	                  \label{chiquadrolensingmismatch} \, ,
\end{eqnarray}
where higher order terms (i.e. $\epsilon^2$) have been neglected.
Equation (\ref{chiquadrolensingmismatch}) shows that a mismatch of order $\epsilon $ in the weak-lensing matrix as described in Eq.~(\ref{barGminusG}) provides no bias in the estimate of $\alpha$ 
but increases its uncertainty by $2 \, \epsilon $. This conclusion of course might change if the mismatch between $\bar G$ and $G$ is more complicated and cannot be modeled as in Eq.~(\ref{barGminusG}).


\section{Discussion and conclusion}
\label{conclusion}

We have studied in detail the D-estimators, which are used to constrain the birefringence angle from CMB observations. 
We have discussed the mathematical framework and pointed out that the close relationship between D-estimators and primordial 
APS can be fruitfully exploited to speed up the analysis. 
In particular if a sufficient large set of simulations are available, then Equations (\ref{DTBp})-(\ref{DEBp}) can be used to build the covariance matrix
through Eqs. (\ref{covTB}), (\ref{covEB}), (\ref{covTBEB}) and (\ref{covEBTB}).
The latter are obtained in the signal plus noise case, but can be generalized to account for residual systematic effects.
We have also computed the covariance matrices in the ideal case, see in Equations (\ref{covTBnew})-(\ref{covEBTBnew})
noting that in the case in which the noise of E and B modes are equal, the covariance matrix for frequentist and Bayesian approach are formally equivalent
and do not depend on $\alpha$. 
We have used this ideal case to describe the behavior of the uncertainties of $\alpha$ when estimated in 
subinterval of the multipoles range in search of a possible scale dependence of the birefringence effect
which can be exploited to disentangle instrumental systematics from physical effects.
Moreover we have shown that the $\chi^2$ formalism built with the D-estimators and the statistics derived from it, is not influenced from weak-lensing effect 
provided that the weak-lensing kernel is exactly known.
When the lensing matrix is misestimated by a term $\epsilon$ proportional to the identity as described in Eq.~(\ref{barGminusG}), then $\alpha$
is not biased, but its uncertainty is increased by $2 \, \epsilon$. 
However, it is not guaranteed that such a $\chi^2$ yields unbiased estimates for $\alpha$ when
the lensing matrix deviates from the real one in a more complicated way with respect to Eq.~(\ref{barGminusG}).


\acknowledgments
A.G, D.M. and P.N. acknowledge support by ASI/INAF Agreement 2014-024-R.1 for the Planck LFI Activity of Phase E2.
%
%
%


\end{document}